\documentclass[3p,times]{elsarticle}
\pdfoutput=1
\usepackage{ecrc}
\usepackage{amsmath}

\volume{00}

\firstpage{1}

\journalname{Physics Procedia}

\runauth{}

\jid{phpro}

\jnltitlelogo{Physics Procedia}

\usepackage{amssymb}

\usepackage[figuresright]{rotating}

\begin{document}
\newcommand{\reactionp}{$\mu^-p\to n\nu$}
\newcommand{\reactiond}{$\mu^-d\to n n \nu$}
\newcommand{\gp}{$g_P^{}$}
\newcommand{\LS}{$\Lambda_S^{}$}
\newcommand{\Ld}{$\Lambda_d^{}$}

\begin{frontmatter}



\dochead{}

\title{Muon capture at PSI}


\author[uw]{Peter Winter\footnote{representing the MuCap \cite{Mucap} and MuSun \cite{Musun} collaborations. Work supported by U.S. National Science Foundation, Department of Energy, NCSA and CRDF, Paul Scherrer Institut and the Russian Academy of Sciences.}}

\address[uw]{Department of Physics, University of Washington, Seattle, WA 98195, USA}

\begin{abstract}
Measuring the rate of muon capture in hydrogen provides one of the most direct ways to study the axial current of the nucleon. The MuCap experiment uses a negative muon beam stopped in a time projection chamber operated with ultra-pure hydrogen gas. Surrounded by a decay electron detector, the lifetime of muons in hydrogen can be measured to determine the singlet capture rate \LS\ to a final precision of 1\%. The capture rate determines the nucleon's pseudoscalar form factor \gp. A first result, \gp$= 7.3 \pm 1.1$, has been published \cite{Andreev:2007wg} and the final analysis of the full statistics will reduce the error by a factor of up to 3. Muon capture on the deuteron probes the weak axial current in the two-nucleon system. Within the framework of effective field theories the calculation of such two-nucleon processes involving the axial current requires the knowledge of one additional low energy constant which can be extracted from the doublet capture rate $\Lambda_d^{}$. The same constant then allows to model-independently calculate related processes such as solar $pp$-fusion or $\nu d$ scattering. The MuSun experiment will deduce $\Lambda_d^{}$ to better than 1.5\%. The experiment uses the MuCap detection setup with a new time projection chamber operated with deuterium at 30\,K and several hardware upgrades. The system is currently fully commissioned and the main physics data taking will start in 2011.
\end{abstract}


\end{frontmatter}


\section{Introduction}
\label{sec_introduction}
A set of two new precision experiments measuring the electroweak processes of muon capture on hydrogen and deuterium were introduced at the Paul Scherrer Institut (PSI) during the last decade. The MuCap experiment measures the singlet rate \LS\ of muon capture on the proton to 1\% precision, while the MuSun experiment aims to deduce the doublet rate \Ld\ for capture on the deuteron to better than 1.5\%. Both measurements determine fundamental constants describing the elctro-weak interaction. The muon capture on the proton will determine the nucleon's pseudo-scalar coupling \gp\ at a final precision of 7\%. This observable describing the one-nucleon axial current has been precisely calculated within chiral pertubation theory (ChPT) \cite{Bernard:1994wn, Kaiser:2003dr} and its measurement at this unprecedented precision will test this low energy QCD result and the underlying chiral symmetry. Similarly to \gp\ for the one-nucleon process, muon capture on the deuteron is determining a low energy constant that arises in treating the two-nucleon axial current within the framework of pion-less effective field theory \cite{Chen:2005ak} or ChPT  \cite{Ando:2001es, Marcucci:2010ts}. Once fixed from a measurement like muon capture on the deuteron, the knowledge of this LEC allows to model-independently calculate related electroweak reactions like solar-pp fusion or $\nu d$ scattering at the same precision as the measured $\mu d$ capture rate.

Both experiments use similar experimental techniques. Muons produced at a proton target are transported in the $\pi$E3 beamline and finally enter a time-projection chamber (TPC) at the center of the experiments. These TPCs are operated with low-density gas; MuCap used 1\% of liquid hydrogen density at room temperature whereas MuSun operates with 30\,K cold deuterium gas at 5\% LH$_2$ density. The chosen conditions are optimal to highly suppress any distorting effects stemming from the formation of muonic molecules (pp$\mu$ and dd$\mu$) and, in case of the deuterium gas, subsequent fusion processes. This technique allows for a selection of muons that are clearly stopped in the gas far away from any high-Z wall materials. A special cleaning system eliminating impurity contaminations in the gas \cite{Ganzha:2007uk} and a special distillation column are crucial elements to guarantee both chemical and isotopic purity of the TPC gas.

A surrounding electron detection system comprising multiwire proportional chambers and a scintillator hodoscope facilitates the full track reconstruction of the decay electron and its timing relative to the muon entering the TPC. A high statistics measurement of more than $10^{10}$ decay electrons allows for a precise measurement of the muon lifetime. The above mentioned capture rates can then be inferred from the difference of the negative muon's lifetime measured in MuCap or MuSun and the positive muon's lifetime \cite{Chitwood:2007pa, Barczyk:2007hp, Webber:2010zf}. In addition, the measurement of the positive muon lifetime in the same chamber gas at similar statistics of $10^{10}$ decays is an important tool for studying various systematic effects.

\section{The MuCap experiment: muon capture on the proton}
\label{sec_mucap}
The elementary electroweak process 
\begin{equation}
\mu^- + p \to n + \nu_{\mu}^{} \label{mucapprocess}
\end{equation}
 is an important probe in understanding the helicity structure of the weak interaction. In the low energy regime, the basic matrix element for this process is reducing to an effective four fermion current-current interaction:
\[
\mathcal{M} \sim G_F^{} V_{ud}^{} \cdot \bar{\psi}_\nu \gamma_\alpha (1-\gamma_5) \psi_\mu \cdot \bar{\psi}_n (V^\alpha_{} - A^\alpha_{}) \psi_p.
\]
Whereas the leptonic current $L^{}_\mu = \bar{\psi}_\nu \gamma_\alpha (1-\gamma_5) \psi_\mu$ is of pure V-A structure, the hadronic current $J_\mu^{} =  \bar{\psi}_n (V^\alpha_{} - A^\alpha_{}) \psi_p$ is dressed due to the substructure of the nucleon. Neglecting second-class currents which must be small due to symmetry principles, the capture process \eqref{mucapprocess} involves $g_V(q_0^2)$, $g_M(q_0^2)$,  $g_A(q_0^2)$, and $g_P(q_0^2)$, i.e. the vector, magnetic, axial, and pseudo-scalar form factors, respectively. The relevant momentum transfer for this process is given by $q_0^2 = -0.88 m_\mu^2$. Since $g_V(q_0^2)$, $g_M(q_0^2)$, and $g_A(q_0^2)$ are well known from other processes, the measurement of \LS\ allows for a determination of $g_P^{}$. This quantity mainly arises from the coupling of the axial current to an intermediate pion. While the dominant pion-pole term has been derived within the partially conserved axial current hypothesis long time ago, it can now be precisely calculated within heavy baryon chiral perturbation theory and its underlying concept of chiral symmetry breaking. Thus, an experimental confirmation of the predicted ChPT result $g_P^{} = 8.26 \pm 0.23$ \cite{Bernard:1994wn, Kaiser:2003dr} is an important test of QCD symmetries.

\begin{figure}[bt]
\centering{\includegraphics[width=0.55\textwidth]{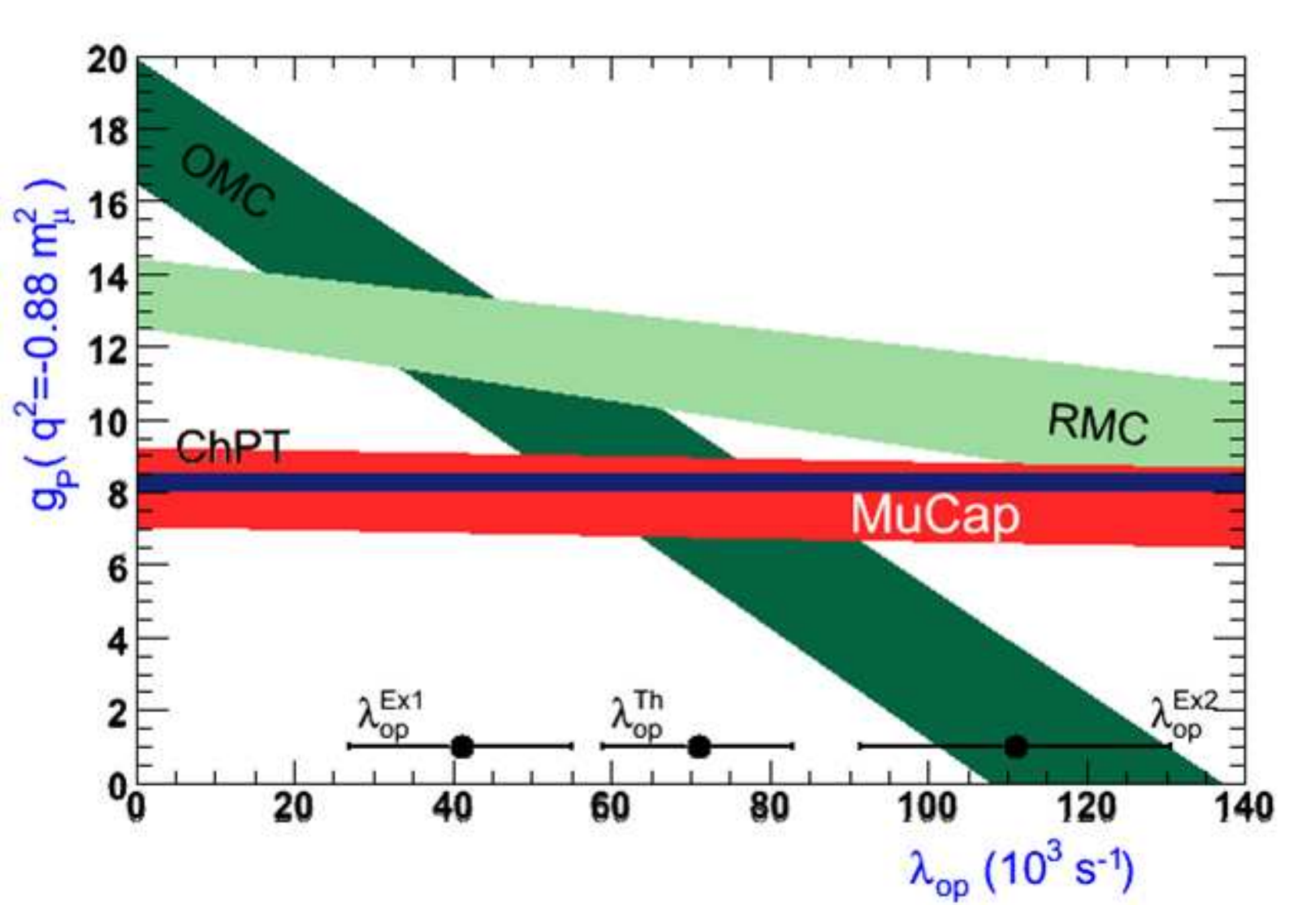}}
\caption{Previous most precise result from OMC \cite{Bardin:1980mi} and RMC \cite{Wright:1998gi} compared to the ChPT prediction \cite{Bernard:1994wn, Kaiser:2003dr} for \gp. Both experiments significantly depend on the poorly known molecular rate ($\lambda_{op}^{Ex1}$ \cite{Bardin:1981cq}, $\lambda_{op}^{Ex2}$ \cite{Clark:2005as}, $\lambda_{op}^{Th}$ \cite{Bakalov:1980fm}). The MuCap result \cite{Andreev:2007wg} does not suffer from this dependence due to the chosen experimental conditions.\label{fig-mucapresults}}
\end{figure}

MuCap measures the singlet rate \LS\ of the ordinary muon capture (OMC) process \eqref{mucapprocess}. The experimental effort on OMC spans a long history and more recently, a first result on the radiative muon capture has been published \cite{Wright:1998gi}. However, as can be seen from figure \ref{fig-mucapresults}, the interpretation of the data was unclear since the most precise previous OMC \cite{Bardin:1980mi} and the RMC result \cite{Wright:1998gi} significantly depend on the molecular ortho-para transition rate $\lambda_{op}$. While details about the various effects from the formation of $pp\mu$ molecules can be found in reviews \cite{Kammel:2010, Gorringe:2002xx}, it is important to note that the experimental conditions of MuCap using a low density gas target at 1\% liquid hydrogen density are essential to overcome the dependence on the poorly known rate $\lambda_{op}$. The first published MuCap result \cite{Andreev:2007wg} shown in the same figure reveals our almost negligible dependence on $\lambda_{op}$ and gives a precise and unambiguous result for \gp\ in agreement with the ChPT prediction.

\begin{figure}
\centering{\includegraphics[width=0.55\textwidth]{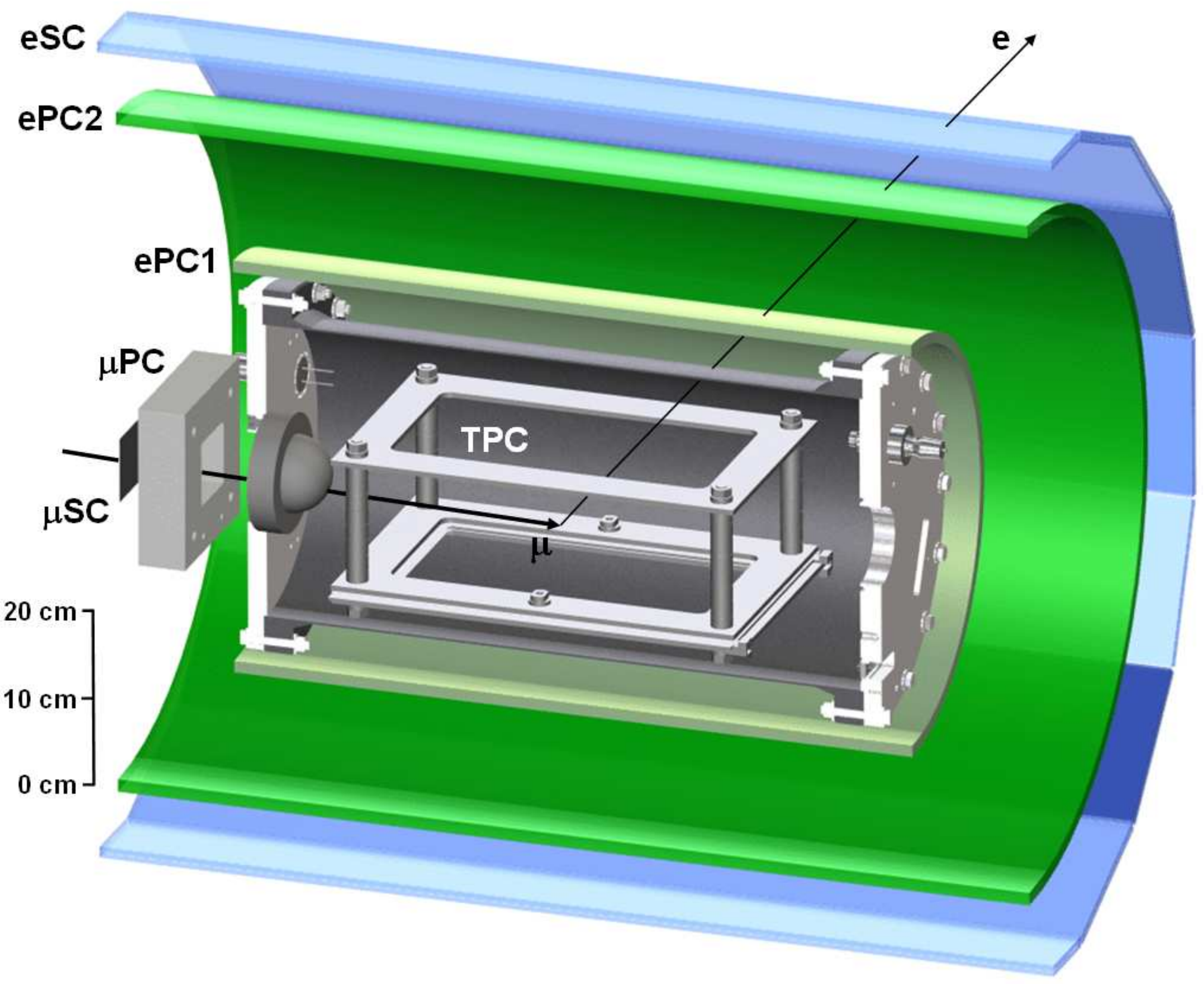}}
\caption{Simplified experimental setup of the MuCap detector.\label{fig-mucap}}
\end{figure}

The MuCap detector as schematically shown in figure \ref{fig-mucap} is installed at the end of the PSI $\pi$E3 beamline at the Paul Scherrer Institut, Switzerland. Muons produced at the proton target are entering the central time-projection chamber where they stop in the hydrogen gas. The muon's arrival is initially registered in the muon entrance scintillator $\mu$SC and a two-plane wire chamber $\mu$PC. Upon registering a muon arrival, an electrostatic kicker deflects the incoming beam to prevent additional muons from entering the target during a 25 microsecond measurement period in the TPC. Pileup of a second muon due to the finite beam extinction is detected in the entrance counters with high efficiency and only leads to a small systematic effect on the lifetime. The TPC is filled with 10 bar ultra-pure hydrogen gas at room temperature which was constantly circulated through an external cleaning unit \cite{Ganzha:2007uk} reducing chemical contaminants to concentrations of less than 10\,ppb. Isotopic purity to better than 6\,ppb was achieved by an initial purification using a cryogenic distillation column that separates isotopes based on the difference in vapor pressure for hydrogen and deuterium. Such high elemental and isotopic purity levels are crucial to reduce any effects from capture on non-hydrogen atoms which can distort the result for the singlet rate \LS. At our concentration levels, these effects are small and well under control. In addition, the experimental setup enables us to monitor and measure the effects stemming from muon capture on chemical impurities as well as the effects from diffusion of $\mu d$ atoms in hydrogen gas.

While the full reconstruction of the muon track by means of the TPC data enables the selection of muon stops far away from any wall material, the decay electrons' trajectories are reconstructed in the cylindrical wire chambers surrounding the TPC. A segmented scintillator hodoscope on the outside provides the fast timing information. The time difference between the electron hodoscope and the muon arrival in the muSC is histogrammed and fit with an exponential and flat background term. Various consistency checks are performed to verify the validity of the fit and several data cuts can be varied to study the stability of the extracted lifetime. Our first result derived from an initial dataset of about $10^9$ decays measured with a first-stage setup has been published \cite{Andreev:2007wg} and is in good agreement with the ChPT prediction. After the data for our published result was taken, the MuCap system underwent some important upgrades like the isotopic purification by means of the deuterium separation column, an improved electronics readout system and the integration of the electrostatic kicker. Since then, we have finished collecting the final statistics of $\sim 1.5\cdot10^{10}$ negative muon decays which will give an improvement in precision by another factor of 3. Currently, the data analysis for the full statistics is at an advanced state and we have finished studying most of the systematic effects. We are planning to unblind our final result in the first half of 2011 which should give \gp\ to the final precision of $\sim$7\%.

\section{The MuSun experiment: muon capture on the deuteron}
\label{sec_musun}

Given the theoretical and experimental accuracy level for the understanding of processes in the one-nucleon sector, it is logical to expand studies on the axial current to the two-nucleon system. This is motivated by the fact that the coupling of the axial current to the two-nucleon systems is related to important astrophysical reactions such as solar-pp fusion \cite{Park:2002yp} or $\nu d$ scattering \cite{Nakamura:2002he} observed by the SNO experiment. In calculations using the framework of effective field theory (EFT), these processes are all described with one additional LEC which absorbs the short distance physics. This LEC is called $L_{1A}$ in pion-less effective field theory \cite{Chen:2005ak} or $\hat{d}_R$ in ChPT \cite{Ando:2001es, Marcucci:2010ts}. The two versions of these EFTs mainly differ in their relevant energy scale which separates the low energy physics from the high energy physics absorbed in the LEC. Once this LEC is precisely known from a process like muon capture on the deuteron, the solar pp-fusion and neutrino scattering can be calculated at a similar precision within the framework of these EFTs.

\begin{figure}
\centering{\includegraphics[width=0.55\textwidth]{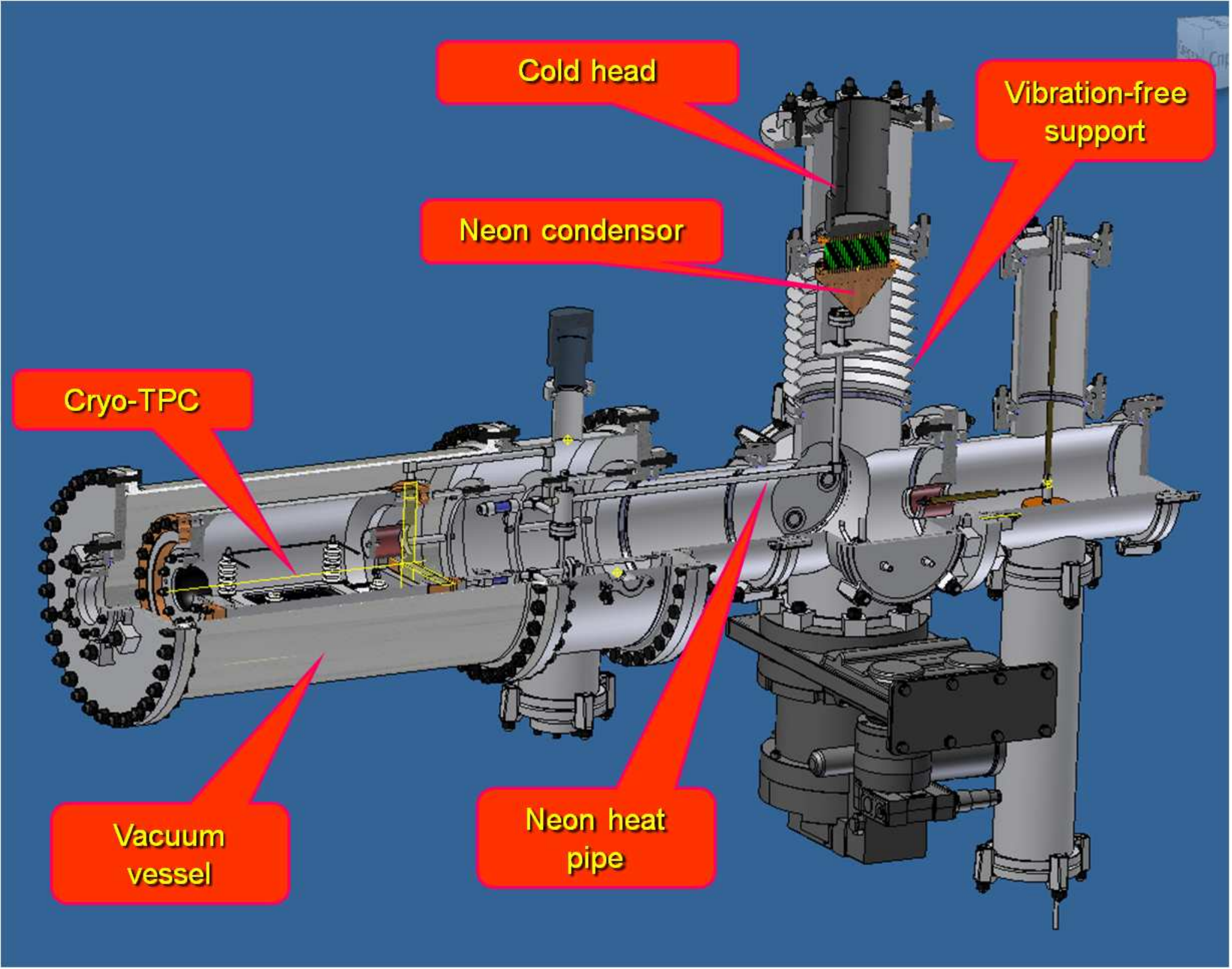}}
\caption{Design of the new cryogenic TPC for the MuSun experiment.\label{fig-cryotpc}}
\end{figure}

The measurement of the doublet capture rate \Ld\ of the process \reactiond\ to a precision of 1.5\% with the MuSun experiment will provide an order of magnitude improvement compared to existing data. The optimal conditions for an unambiguous interpretation of the measured rate with respect to the $\mu d$ hyperfine and molecular $dd \mu$ states are at low temperature (30 K) and a gas density of 5\% of liquid hydrogen. Therefore, the collaboration has built and commissioned a new cryogenic TPC operated with high-purity deuterium. The schematic design of the vibration-free neon cooling system and the new TPC is shown in Fig. \ref{fig-cryotpc}. As for the rest of the experimental setup, we use the same detectors as for the MuCap experiment.

Since the \reactiond\ leads to additional charged ejectiles via the fast $dd \mu$ fusion channels, a good energy resolution in the TPC needs to be achieved. For that, the new TPC is operated in an ionization mode without gas gain. In addition, we have replaced the entire readout system with new improved low-noise amplifiers and a readout of the TPC signal waveforms with flash-ADCs. This will allow separation of the initial muon ionization charge from charged fusion signals which is important to eliminate any time-dependent distortions to the measured capture rate.

At the moment, the collaboration has achieved stable conditions at 30\,K and 80\,kV drift voltage with the new cryogenic TPC. The chemical purity of the gas is excellent since we use the same cleaning system as for MuCap. We will isotopically purify the deuterium gas using the same distillation column as MuCap, whereby a protium depletion of $10^{-4}$ is sufficient to avoid the necessity of systematic corrections to our measured rate. Right now, we are commissioning the full system and we have already shown that our energy resolution with the new readout system is close to our final aim. Therefore, we are already collecting valuable data to understand the necessary improvements required before we can start to collect the $\sim$$10^{10}$ $\mu^-$ decays in deuterium needed to obtain the proposed precision on \Ld. The main data taking is scheduled to start in 2011.





\bibliographystyle{elsarticle-num}
\bibliography{winter}







\end{document}